\documentclass[slac_one]{revtex4}
\usepackage{graphicx}
\usepackage{fancyhdr}
\begin{document}

\title{{\small{2005 International Linear Collider Workshop - Stanford, U.S.A.}}\\
\vspace{12pt}
Resonant \boldmath{$H$ and $A$} Mixing in CP--noninvariant 2HDM and MSSM}

\author{S.Y. Choi}
\affiliation{Department of Physics, Chonbuk National University,
             Chonju 561--756, Korea}

\author{J. Kalinowski}
\affiliation{Institute of Theoretical Physics, Warsaw University,
PL--00681,
             Warsaw, Poland}

\begin{abstract}
In the general two--Higgs--doublet model, including the minimal
supersymmetric standard model (MSSM) as its specific realization,
the two heavy neutral Higgs bosons are nearly degenerate in the
decoupling limit. If the theory is CP--noninvariant, the mixing
between the heavy states can strongly be affected by the decay
widths. We develop the formalism describing this CP--violating
non--hermitian mixing and provide some interesting experimental
signatures of the CP--violating mixing at a $\gamma\gamma$
collider with polarized beams in the context of the
CP--noninvariant MSSM.
\end{abstract}

\maketitle

\thispagestyle{fancy}


\section{INTRODUCTION}

The MSSM is a specific realization of general scenarios that
include two doublet fields in the Higgs sector. After three
Goldstone fields are absorbed by electroweak gauge bosons, the
remaining five fields give rise to physical states. At the tree
level, the MSSM Higgs sector is CP invariant, with two CP--even
and one CP--odd neutral states. However, the MSSM offers new
sources of CP violation, which render the Higgs sector
CP--noninvariant at the loop level. In such CP--noninvariant
theories the three neutral states mix to form a triplet with both
even and odd components in the wave--functions under CP
transformations \cite{kalinowski_gunion, pilaftsis, pilaftsis_wagner}.
The mixing can become very large if the states are nearly
mass--degenerate. This situation is naturally realized for
supersymmetric theories in the decoupling limit
\cite{cp_invariant_decoupling} in which two of the neutral states
are heavy.

In the present report we describe a simple quantum mechanical (QM)
formalism for the CP--violating resonant $H/A$ mixing in the
decoupling limit and then discuss some experimental signatures of
the CP--violating mixing in Higgs production and decay processes
at a photon collider with polarized photon beams.

\section{MIXING FORMALISM}

The self-interaction of two Higgs doublets in a CP--noninvariant
theory is
generally described by the potential \cite{cp_invariant_decoupling}\\[-6mm]
\begin{eqnarray}
{\cal V}&=&
m_{11}^2\Phi_1^\dagger\Phi_1+m_{22}^2\Phi_2^\dagger\Phi_2
           -[m_{12}^2\Phi_1^\dagger\Phi_2+{\rm h.c.}]
           +\frac{1}{2}\lambda_1(\Phi_1^\dagger\Phi_1)^2
           +\frac{1}{2}\lambda_2(\Phi_2^\dagger\Phi_2)^2\nonumber\\[-1mm]
        && +\lambda_3(\Phi_1^\dagger\Phi_1)(\Phi_2^\dagger\Phi_2)
           +\lambda_4(\Phi_1^\dagger\Phi_2)(\Phi_2^\dagger\Phi_1)
           +\left\{\frac{1}{2}\lambda_5(\Phi_1^\dagger\Phi_2)^2
           +\big[\lambda_6(\Phi_1^\dagger\Phi_1)
                +\lambda_7(\Phi_2^\dagger\Phi_2)\big]\,
                \Phi_1^\dagger\Phi_2+{\rm h.c.}\right\}
\label{eq:higgs_potential}
\end{eqnarray}
where $\Phi_{1,2}$ denote two complex $Y=1$, SU(2)$_L$ iso-doublet
scalar fields. The coefficients are in general all non--zero. The
parameters $m^2_{12}, \lambda_{5,6,7}$ can be complex,
incorporating the CP-noninvariant elements in the interactions.
The neutral components of the scalar fields $\Phi_1$ and $\Phi_2$
are assumed to develop non--zero vacuum expectation values (vevs)
$\langle \phi^0_1\rangle= v_1/\sqrt{2}$ and $\langle
\phi^0_2\rangle =v_2/\sqrt{2}$, which can be chosen real and
positive without loss of generality. As usual,
$v=(v_1^2+v_2^2)^{1/2} = 246$ GeV.

It is useful to rotate the Higgs fields $\Phi_{1,2}$ to the
$\Phi_{a,b}$ basis with the angle $\beta$ satisfying
$\tan\beta=v_2/v_1$ (we exploit the abbreviations
$t_{\beta}=\tan\beta$, $c_\beta=\cos\beta$, $s_{2\beta} = \sin
2\beta$ etc.). In this basis only the field $\Phi_a$ develops a
non--zero vev
\begin{eqnarray}
\Phi_a = \left(\begin{array}{c}
                G^+ \\
             \frac{1}{\sqrt{2}}
             \left(v+H_a+i G^0\right)
               \end{array}\right),
 \qquad
\Phi_b= \left(\begin{array}{c}
                H^+ \\
             \frac{1}{\sqrt{2}}
             \left(H_b+i A\right)
              \end{array}\right)
\label{eq:ab_basis}
\end{eqnarray}
and the three fields $G^{\pm,0}$ can be identified as the would-be
Goldstone bosons, while $H^{\pm}, H_{a,b}$ and $A$ give rise to
physical Higgs bosons. The real mass matrix ${\cal M}^2_{0}$ of
neutral Higgs fields in the basis of $H_a, H_b, A$, which is
hermitian and symmetric by CPT invariance, can easily be derived
from the Higgs potential (\ref{eq:higgs_potential}) after the
rotations:
\begin{eqnarray}
{\cal M}^2_{0R}=v^2 \left(\begin{array}{ccc}
       \lambda         &    -\hat{\lambda}             & -\hat{\lambda}_p \\[0mm]
      -\hat{\lambda}   & \lambda-\lambda_A +M^2_A/v^2  & -\lambda_p  \\[0mm]
      -\hat{\lambda}_p &   -\lambda_p                  &  M^2_A/v^2
                   \end{array}\right)
\label{eq:mass_matrix_ab}
\end{eqnarray}
after eliminating $m^2_{11,22}, m^{2I}_{12}$ from the minimization
conditions, and exchanging $m^{2R}_{12}$ for the {\it auxiliary}
parameter $M^2_A$. It is defined by the relation
\begin{eqnarray}
m^{2R}_{12} = {\textstyle\frac{1}{2}} ( M^2_A s_{2\beta}
            + v^2(\lambda^R_5 s_{2\beta} + \lambda^R_6 c^2_\beta
                  +\lambda^R_7 s^2_\beta) )
\end{eqnarray}
and it will be one of the key parameters in the system.
The $\lambda, \hat{\lambda}$ and $\lambda_A$ parameters are
functions of the real parts, while $\lambda_p$ and
$\hat{\lambda}_p$ are functions of the imaginary parts of the
parameters $\lambda_i$ in Eq.~(\ref{eq:higgs_potential}); their
explicit form can be found in Ref.~\cite{CKLZ}.

In a CP--invariant theory all $\lambda_i$ couplings are real and
the off--diagonal elements $\lambda_p, \hat{\lambda}_p$ vanish.
Thus the neutral mass matrix breaks into the CP--even $2\times 2$
part, and the [stand--alone] CP--odd part. The $2\times 2$ part
gives rise to two CP--even neutral mass eigenstates $h$ and $H$,
while $M_A$ is identified as the mass of the CP--odd Higgs boson
$A$. In the CP--violating case, however, all three states mix
leading to $H_{1,2,3}$ mass eigenstates with no definite CP
parities.

For small mass differences, the mixing of the states is strongly
affected by their widths. This is a well--known phenomenon for
resonant mixing \cite{resonant_mixing_zerwas} and has also been
recognized for the Higgs sector \cite{resonant_cp_pilaftsis}. The
hermitian mass matrix (\ref{eq:mass_matrix_ab}) has therefore to
be supplemented by the anti--hermitian part $-iM\Gamma$
incorporating the decay matrix \cite{ww_form}
\begin{eqnarray}
{\mathcal M}^2 = {\mathcal M}^2_0 - i M \Gamma
\end{eqnarray}
This matrix includes the widths of the Higgs states in the
diagonal elements as well as the transition elements within any
combination of pairs. They are particularly important in the case
of nearly mass--degenerate states. All these elements
$(M\Gamma)^{AB}_{ab}$ are built up by loops of the fields $(AB)$
in the self-energy matrix $\langle h_a h_b\rangle$ of the Higgs
fields.

In general, the light Higgs boson, the fermions and electroweak
gauge bosons, and in supersymmetric theories, gauginos, higgsinos
and scalar states may contribute to the loops in the propagator
matrix. In the physically interesting case of decoupling, the
mixing structure simplifies considerably allowing for a very
simple and transparent approach \cite{CKLZ}. Alternative approach
requires a full coupled--channel analysis \cite{coupled_channel}.

\subsection{Decoupling Limit}

The decoupling limit \cite{cp_invariant_decoupling} is defined by
the inequality $M^2_A\,\, \gg\,\, |\lambda_i|\, v^2$ with
$|\lambda_i| \lesssim O(1)$. In this limit the $H_a$ state becomes
the CP--even light Higgs boson $h$ and decouples from $H_b$ and
$A$. The heavy states $H=H_b$ and $A$ are nearly mass degenerate,
which turns out to be crucial for large mixing effects between $H$
and $A$. It is therefore enough to consider a lower--right
$2\times 2$ submatrix of the matrix (\ref{eq:mass_matrix_ab}) for
the heavy $H/A$ states which we write as follows
\begin{eqnarray}
{\mathcal M}^2_{HA}
  = \left(\begin{array}{cc}
   M^2_H - i M_H \Gamma_H  & \Delta^2_{HA}  \\[0mm]
    \Delta^2_{HA}        & M^2_A - i M_A \Gamma_A
    \end{array}\right)
\label{eq:complex_mass_matrix_squared}
\end{eqnarray}
where $\Delta^2_{HA}$ also consists of a real dissipative part and an
imaginary absorptive part. Moreover, the couplings of the heavy
Higgs bosons to gauge bosons and their supersymmetric partners are
suppressed. In the case of all supersymmetric particle
contributions to be suppressed either by couplings or by phase
space in $M\Gamma$, it is sufficient to consider only loops built
up by the light Higgs boson and top quark; for the explicit form of
the light Higgs boson and top quark loop contributions to the matrix
$M\Gamma$, we refer to Ref.~\cite{CKLZ}. The loops also contribute to
the real part of the mass matrix, either renormalizing the $\lambda$
parameters of the Higgs potential or generating such parameters if
not present yet at the tree level.

\subsection{Physical Masses and States}

The symmetric complex mass--squared matrix ${\cal M}^2$ in
Eq.(\ref{eq:complex_mass_matrix_squared}) can be diagonalized
through a {\it complex rotation}
\begin{eqnarray}
{\mathcal M}^2_{H_i H_j}
  = \left(\begin{array}{cc}
   M^2_{H_2} - i M_{H_2} \Gamma_{H_2}  & 0                  \\[0mm]
      0      & M^2_{H_3} - i M_{H_3} \Gamma_{H_3}
    \end{array}\right)
  =  C {\mathcal M}^2_{HA} C^{-1}
\label{eq:complex_massij_matrix_squared}
\end{eqnarray}
where the mixing matrix and the mixing angle are given by
\begin{eqnarray}
C = \left(\begin{array}{cc}
      \cos\theta  & \sin\theta \\[0mm]
     -\sin\theta  & \cos\theta
    \end{array}\right), \qquad
X = \frac{1}{2}\tan 2\theta
  = \frac{\Delta^2_{HA}}{M^2_H - M^2_A
                          -i\left[M_H \Gamma_H - M_A\Gamma_A\right]}
\end{eqnarray}
A non--vanishing (complex) mixing parameter $X\neq 0$ requires
CP--violating transitions between $H$ and $A$ either in the real
mass matrix, $\lambda_p\neq 0$, or in the decay mass matrix,
$(M\Gamma)_{HA}\neq 0$, [or both]. However, note that even for
nearly degenerate masses, the mixing could be suppressed if the
widths were significantly different. As a result, the mixing
phenomena are strongly affected by the form of the decay matrix
$M\Gamma$. Since the difference of the widths enters through the
denominator in $X$, the modulus $|X|$ becomes large for small
differences and small widths.

The mixing shifts the Higgs masses and widths in a characteristic
pattern \cite{resonant_mixing_zerwas}. The two complex mass values
after and before diagonalization are related by the complex mixing
angle $\theta$:
\begin{eqnarray}
M^2_{H_3} - M^2_{H_2} - i\left(M_{H_3} \Gamma_{H_3} - M_{H_2}
\Gamma_{H_2}\right)
 =\left[M^2_A-M^2_H-i (M_A \Gamma-M_H \Gamma_H)\right]
           \times \sqrt{1+4X^2}
\end{eqnarray}
Since the eigenstates of the complex, non--hermitian matrix ${\cal
M}^2$ are no longer orthogonal, the ket and bra mass eigenstates
have to be defined separately: $|H_i \rangle = C_{i\alpha}
|H_\alpha\rangle$ and $\langle\widetilde{H}_i| =
C_{i\alpha}\langle H_\alpha|$ ($i=2, 3$ and $H_\alpha=H,A$).
The final state $F$ in heavy Higgs formation from the initial
state $I$ is, then, described with the amplitude
\begin{eqnarray}
\langle F|H|I\rangle = \sum_{i=2,3}\, \langle F| H_i \rangle\,
                       \frac{1}{s-M^2_{H_i}+i M_{H_i} \Gamma_{H_i}}\,
                       \langle\, \widetilde{\!H}_i|I\rangle
\end{eqnarray}
where the sum runs only over diagonal transitions in the
mass--eigenstate basis.

\section{EXPERIMENTAL SIGNATURES}

To illustrate the general QM results in a realistic example, we
adopt a specific MSSM scenario with the source of CP--violation
localized in the complex trilinear coupling $A_t$ of the soft
supersymmetry breaking part involving the top
squark.\footnote{This assignment is compatible with the bounds on
CP--violating SUSY phases from experiments on electric dipole
moments \cite{edm}.} All other interactions are assumed to be
CP--conserving. For $\phi_A\neq0,\,\pi$, the stop--loop
corrections induce the CP--violation  in the effective Higgs
potential (\ref{eq:higgs_potential}). The effective $\lambda_i$
parameters have been calculated in Ref.~\cite{pilaftsis_wagner} to
two--loop accuracy; to illustrate the crucial points we take the
dominant one--loop $t/\tilde t$ contributions.

More specifically, we take a typical set of parameters from
Ref.~\cite{cpsuperh},
\begin{eqnarray}
M_S = 0.5\,\, {\rm TeV},\quad |A_t|= 1.0\,\, {\rm TeV},\quad \mu=
1.0\,\, {\rm TeV}; \quad \tan\beta=5 \label{eq:parameters}
\end{eqnarray}
and change the phase $\phi_A$ of the trilinear parameter $A_t$.
With $\phi_A=0$ we find the following values of the light and
heavy Higgs masses and decay widths, and the stop masses:
\begin{eqnarray}
\begin{array}{lll}
M_h = 129.6\, {\rm GeV},\,  M_H = 500.3\, {\rm GeV},\, M_A =
500.0\, {\rm GeV};\ \ \Gamma_H = 1.2\, {\rm GeV},\, \Gamma_A =
1.5\, {\rm GeV};\ \ m_{\tilde{t}_{1/2}}= 372/647\, {\rm  GeV}
\end{array}
\end{eqnarray}
Clearly, with the mass splitting of 0.3 GeV, the heavy Higgs
states are not distinguishable.
\begin{figure}[t]
\begin{center}
\vskip 0.2cm
\includegraphics[width=4.3cm,height=4.5cm,clip=true]{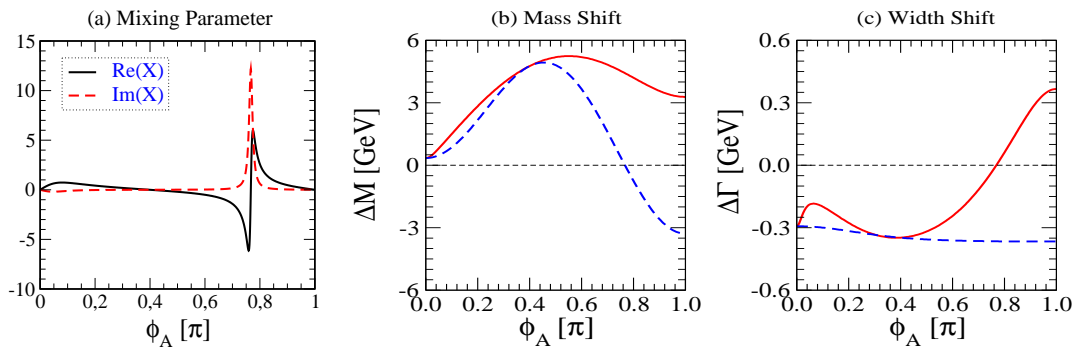}~~
\includegraphics[width=9.6cm,height=4.5cm]{shift_mssm.eps}
\end{center}
\vskip -0.7cm \caption{The $\phi_A$ dependence of (a)
     the mixing parameter $X$ and of the shifts of (b) masses and (c)
     widths with the phase $\phi_A$ evolving from $0$ to $\pi$ for
     $\tan\beta=5$, $M_A=0.5$ TeV and couplings as specified in
     the text; in (b,c) the mass and width differences
     without mixing  are shown by the broken lines.
    $\Re{\rm e}/\Im{\rm m} X(2\pi-\phi_A)
     =+\Re{\rm e}/\!\!-\!\Im{\rm m} X(\phi_A)$ for angles above $\pi$.}
\label{fig:ReIm-mssm}
\end{figure}
When the phase $\phi_A$ is turned on,\footnote{With one phase
$\phi_A$, the complex mixing parameter
  $X$ obeys the relation $X(2\pi-\phi_A)=X^*(\phi_A)$, implying  all CP--even
  quantities
  symmetric and all CP--odd quantities anti--symmetric about $\pi$.}
the CP composition, the masses and the decay widths of heavy
states are strongly affected, as shown in
Figs.~\ref{fig:ReIm-mssm}(a), (b) and (c), while the mass of the
light Higgs boson $h$ is not. The heavy two--state system shows a
very sharp resonant CP--violating mixing, purely imaginary a
little above $\phi_A = 3\pi/4$, Fig.~\ref{fig:ReIm-mssm}(a). The
mass shift is enhanced by more than an order of magnitude if the
CP--violating phase rises to non-zero values, reaching a maximal
value of $\sim 5.3$ GeV; the width shift changes   between $-0.3$
and $+0.4$ GeV. As a result, the two mass--eigenstates should
become clearly distinguishable at future colliders, in particular
at a photon collider \cite{Muhlleitner:2001kw}. Moreover, both
states have significant  admixtures of CP--even and CP--odd
components in the wave--functions. Since  $\gamma\gamma$ colliders
offer unique conditions for probing the CP--mixing
\cite{general_method,gg_h,h_tt}, we discuss two experimental
examples: (a) Higgs formation in polarized $\gamma\gamma$
collisions and (b) polarization of top quarks in Higgs decays,
where  spectacular signatures of resonant mixing can be expected.

{\bf (a)} The amplitude of the reaction $\gamma\gamma\rightarrow
H_i\rightarrow F$ is a superposition of $H_2$ and $H_3$ exchanges.
For equal helicities $\lambda=\pm 1$ of the two photons, the
amplitude reads
\begin{equation}
{\cal M}^F_{\lambda}
  =\sum_{i=2,3}\, \langle F| H_i\rangle\, \frac{1}{s-M^2_{H_i}+i
  M_{H_i}\Gamma_{H_i}}
  \left[ S^\gamma_i (s)+i\lambda P^\gamma_i(s) \right]
\end{equation}
where $\sqrt{s}$ is the $\gamma\gamma$ energy and the
loop--induced $\gamma\gamma H_i$ scalar and pseudoscalar form
factors, $S^\gamma_i(s)$ and $P^\gamma_i(s)$, are related to the
well--known conventional $\gamma\gamma H/A$ form factors,
$S^\gamma_{H,A}$ and $P^\gamma_{H,A}$. For their relation and
explicit form we refer to Refs.~\cite{CKLZ} and \cite{cpsuperh}.
In our scenario the Higgs--$tt$ couplings are assumed to be
CP--conserving, implying negligible top-loop contributions to
$P^\gamma_H$ and $S^\gamma_A$ since the gluino mass is
sufficiently heavy compared with the stop masses, while the
$\tilde{t}_1$ loop generates a non-negligible CP-violating
amplitude $S^{\gamma}_A$. In the region of strong mixing on which
we focus, however, the CP-violating vertex corrections have only a
small effect on the experimental asymmetries compared with the
large impact of CP-violating Higgs-boson mixing.

Polarized photons provide a very powerful tool to investigate the
CP properties of Higgs bosons. With linearly polarized photons one
can project out the CP--even and CP--odd components of the $H_i$
wave--functions by arranging the photon polarization vectors to be
parallel or perpendicular. On the other hand, circular
polarization provides us with a direct insight into  the
CP-violating nature of Higgs bosons. Two asymmetries are of
interest
\begin{eqnarray}
{\cal A}_{lin}
  = \frac{\sigma_\parallel- \sigma_\perp}{\sigma_\parallel+ \sigma_\perp},
\hspace{1.3cm} {\cal A}_{hel}
  =\frac{\sigma_{++} -\sigma_{--}}{\sigma_{++}+ \sigma_{--}}
\end{eqnarray}
where $\sigma_\parallel$, $\sigma_\perp$ and $\sigma_{++}$,
$\sigma_{--}$ are the corresponding total $\gamma\gamma$ fusion
cross sections for linear and circular polarizations,
respectively. Though  CP--even, the asymmetry ${\cal A}_{lin}$ can
serve as a powerful tool nevertheless to probe CP--violating
admixtures to the Higgs states since $|{\cal A}_{lin}|<1$ requires
both $S^\gamma_i$ and $P^\gamma_i$ non-zero couplings. A more
direct probe  of CP--violation due to $H/A$ mixing is provided by
the CP--odd (and also CP$\tilde{\rm T}$--odd) asymmetry ${\cal
A}_{lin}$.
\begin{figure}[htb]
\vskip -0.1cm
\includegraphics[width=9.9cm,height=4.5cm,clip=true]{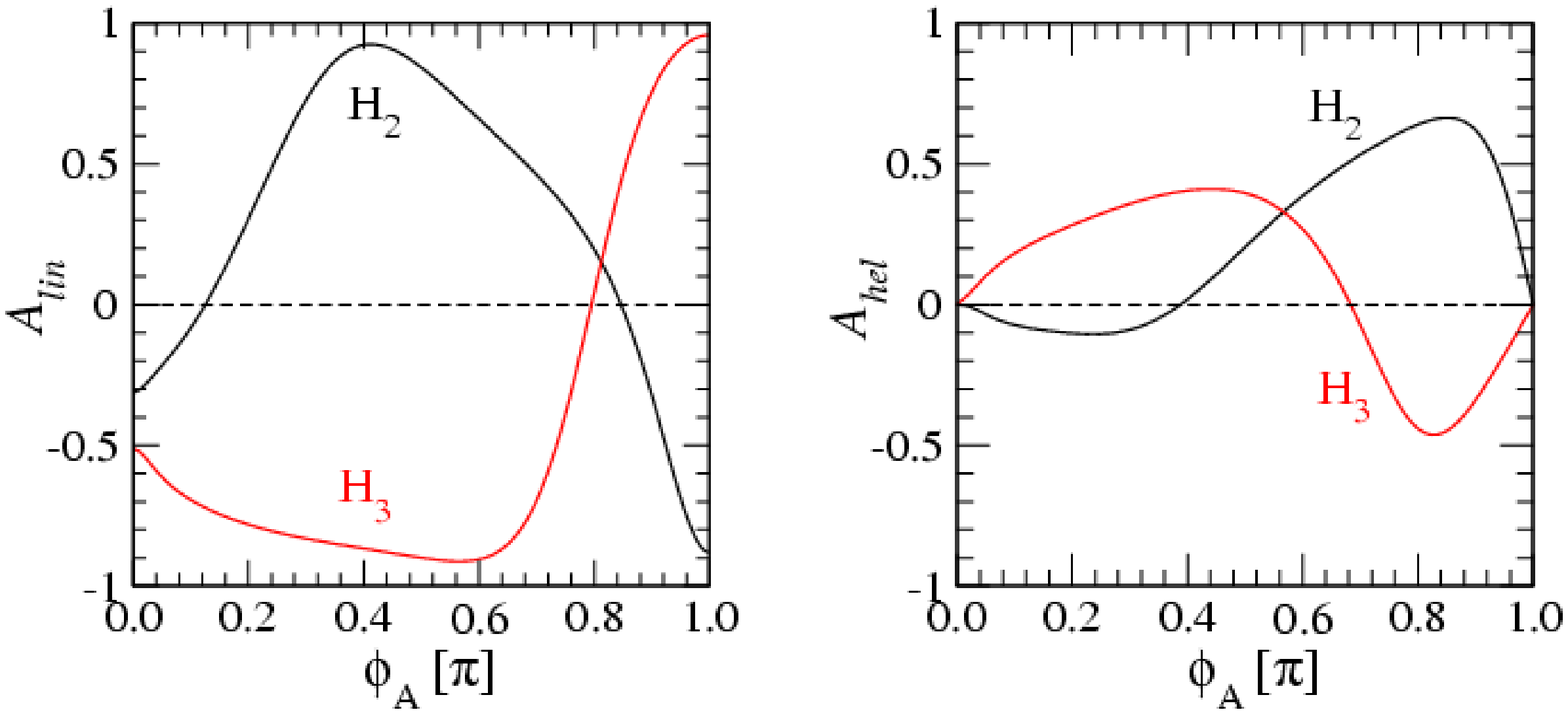} ~~ ~
\includegraphics[width=4.7cm,height=4.5cm,clip=true]{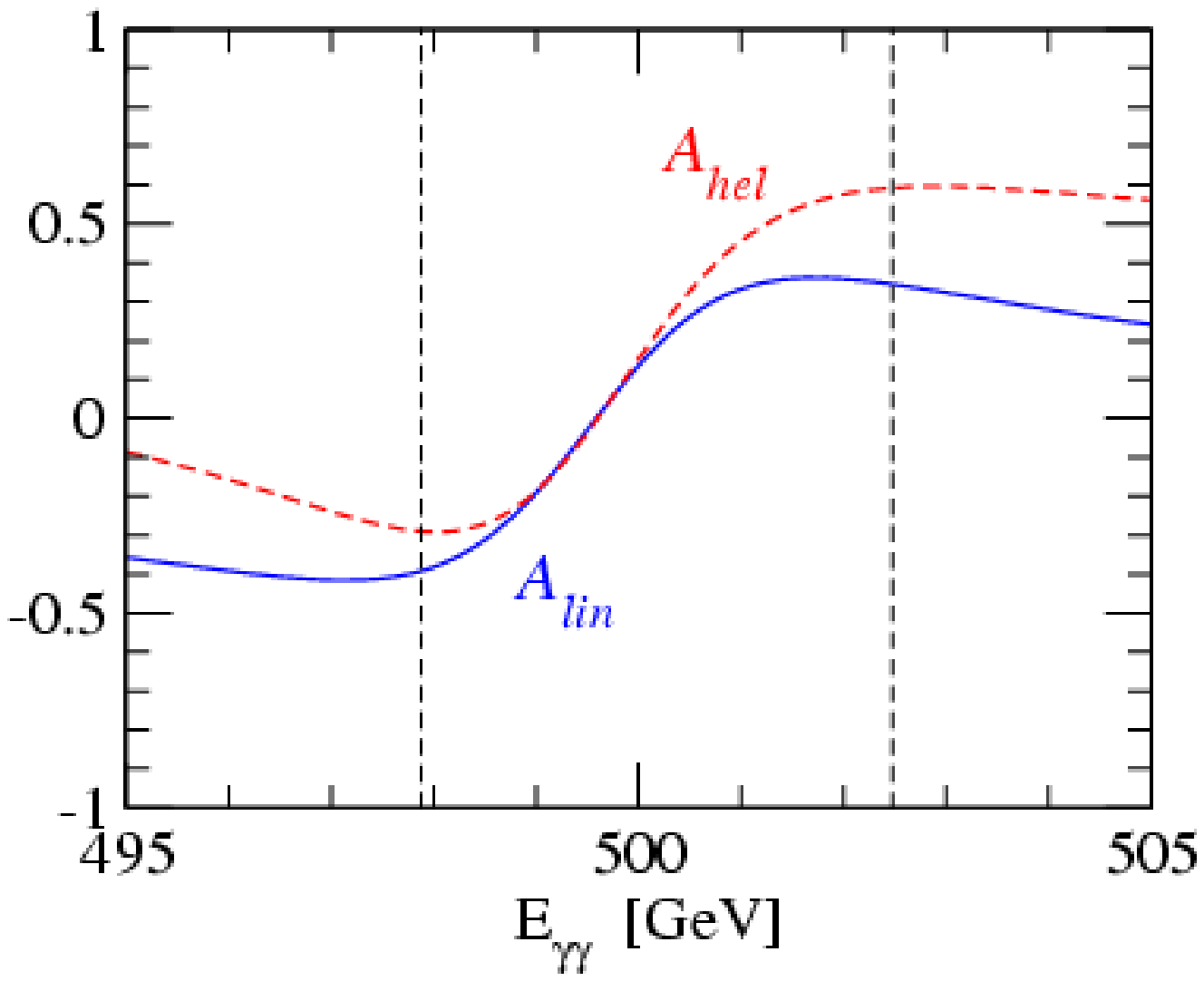}
\vskip -0.2cm \caption{The $\phi_A$ dependence of the CP--even and
CP--odd correlators,
         ${\cal A}_{lin}$ (left panel) and ${\cal A}_{hel}$
         (center
         panel), at the poles of $H_2$ and $H_3$, respectively, and the
         $\gamma\gamma$ energy dependence (right panel) of the correlators,
         ${\cal A}_{lin,hel}$,  for $\phi_A=3\pi/4$ in the production process
         $\gamma\gamma \rightarrow H_i$ in the limit in which $H/A$
         mixing is the dominant CP--violating effect.
         The same parameter set as in Fig.~\ref{fig:ReIm-mssm} is employed.
         The vertical lines on the right panel mark positions of the two mass
         eigenvalues, $M_{H_3}$ and $M_{H_2}$.}
\label{fig:asym_rrh}
\end{figure}

Fig.~\ref{fig:asym_rrh} show the $\phi_A$ dependence of the
asymmetries ${\cal A}_{lin}$ and ${\cal A}_{hel}$ at the poles of
$H_2$ and of $H_3$, respectively, for the same parameter set as in
Fig.~\ref{fig:ReIm-mssm} and with the common SUSY scale
$M_{\tilde{Q}_3}=M_{\tilde{t}_R}= M_S=0.5$ TeV for the soft SUSY
breaking top squark mass parameters. By varying the $\gamma\gamma$
energy from below $M_{H_3}$ to above $M_{H_2}$, the asymmetries,
${\cal A}_{lin}$ (blue solid line) and ${\cal A}_{hel}$ (red
dashed line), vary from $-0.39$ to $0.34$ and from $-0.29$ to
$0.59$, respectively, as demonstrated on the right panel of
Fig.~\ref{fig:asym_rrh} with $\phi_A=3\pi/4$, a phase value close
to resonant CP--mixing.

{\bf (b)} A second observable of interest  is the polarization of
the top quarks in $H_i$ decays produced by $\gamma\gamma$ fusion
or elsewhere in various production processes at an $e^+e^-$ linear
collider and LHC $H_{2,3} \rightarrow t\bar{t}$. Even if the $H/A
tt$ couplings are [approximately] CP--conserving, the complex
rotation matrix $C$ may mix the CP--even $H$ and CP--odd $A$
states leading to CP--violation. In the production--decay process
$\gamma\gamma\rightarrow H_i\rightarrow t\bar{t}$, two CP--even
and CP--odd correlators between the transverse $t$ and $\bar{t}$
polarization vectors $s_{\bot},\bar{s}_{\bot}$
\begin{eqnarray}
{\cal C}_\parallel = \left\langle s_\perp \cdot \bar{s}_\perp
                 \right\rangle\qquad {\rm and},\qquad
{\cal C}_\perp = \left\langle \hat{p}_t\cdot
(s_\perp\times\bar{s}_\perp)
                 \right\rangle
\end{eqnarray}
can be extracted from the azimuthal--angle correlation between the
two decay planes $t\rightarrow bW^+$ and $\bar{t}\rightarrow
\bar{b}W^-$ \cite{general_method}.

\begin{figure}[htb]
\vskip -0.1cm
\includegraphics[width=9.9cm,height=4.5cm,clip=true]{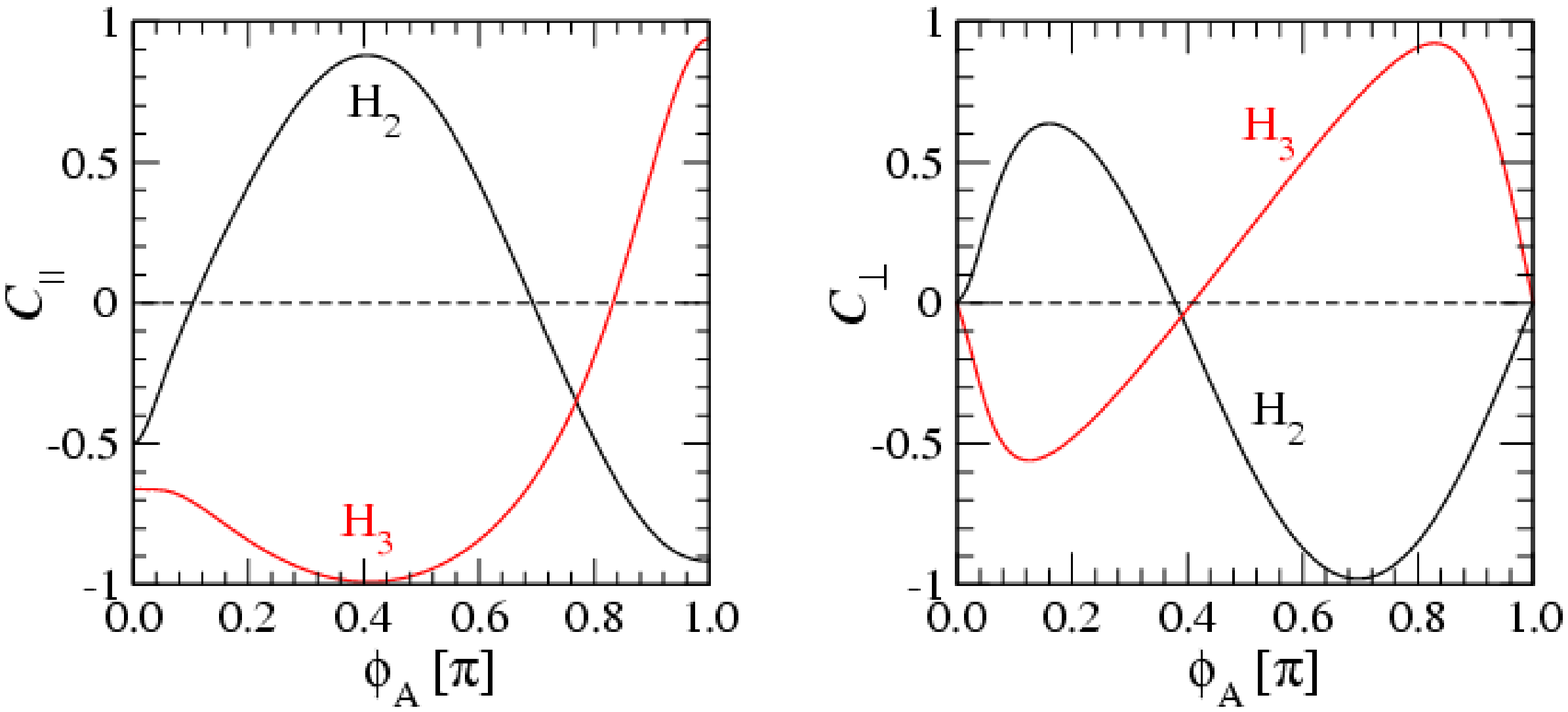} ~~ ~
\includegraphics[width=4.7cm,height=4.5cm,clip=true]{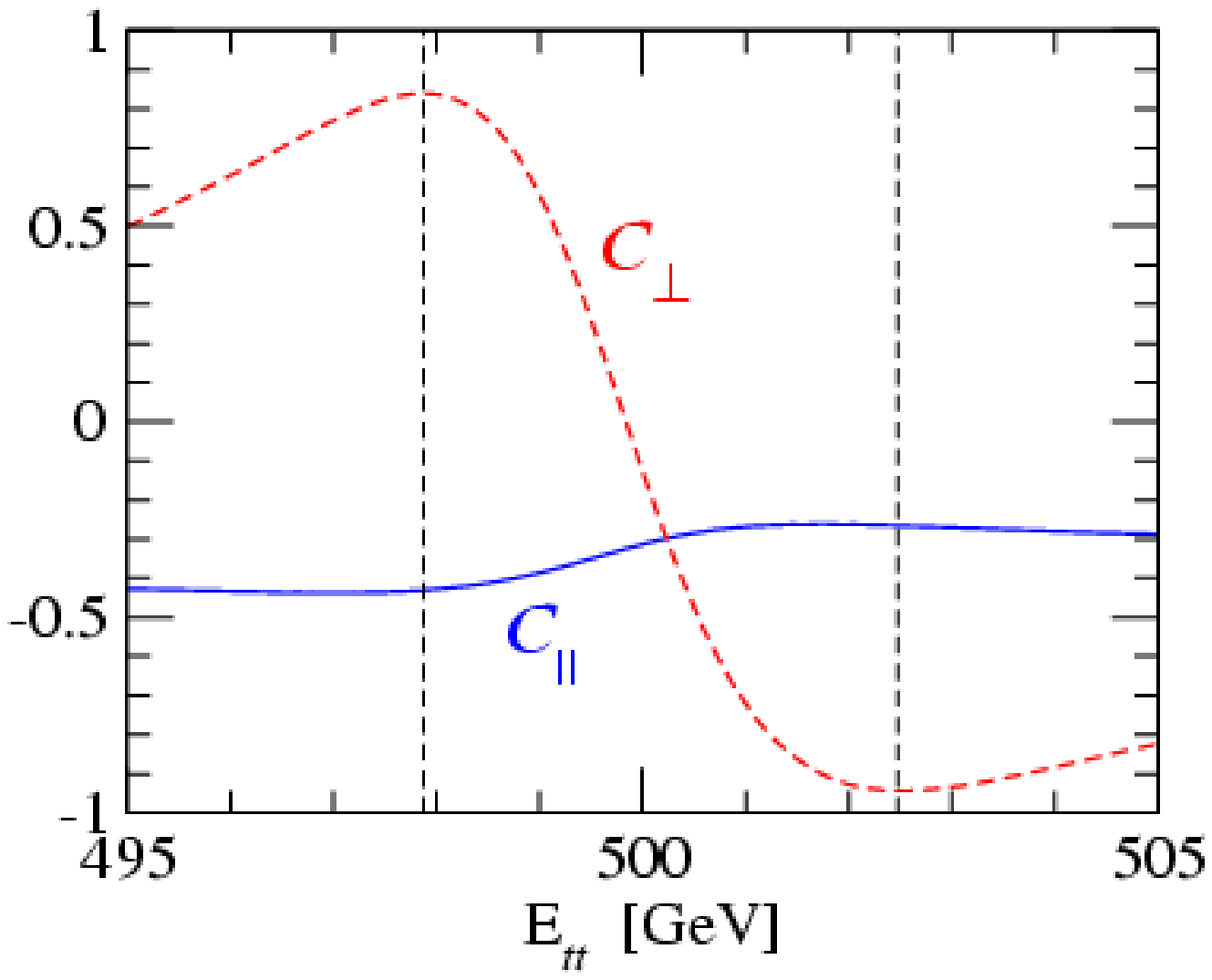}
\vskip -0.2cm \caption{The $\phi_A$ dependence of the CP--even and
CP--odd correlators,
         ${\cal C}_\parallel$ (left panel)  and ${\cal C}_\perp$
         (center panel), at the pole of $H_2$ and $H_3$ and the invariant
         $t\bar{t}$ energy dependence (right panel) of the correlators
         ${\cal C}_{\parallel,\perp}$ for $\phi_A=3\pi/4$ in the
         production--decay chain $\gamma\gamma \rightarrow H_i\rightarrow
         t\bar{t}$. [Same SUSY parameter set as in Fig.\ref{fig:asym_rrh}.]}
\label{fig:asym_htt}
\end{figure}

Fig.~\ref{fig:asym_htt} shows the $\phi_A$ dependence of the
CP--even and CP--odd asymmetries, ${\cal C}_\parallel$ and ${\cal
C}_\perp$, at the poles of $H_2$ and of $H_3$, left and center
panels respectively. If the invariant $t\bar{t}$ energy is varied
throughout the resonance region, the correlators ${\cal
C}_\parallel$ (blue solid line) and ${\cal C}_\perp$ (red dashed
line) vary characteristically from $-0.43$ to $-0.27$
[non--uniformly] and from $0.84$ to $-0.94$, respectively, as
shown in the right panel of Fig.~\ref{fig:asym_htt}.

\section{CONCLUSIONS}

Exciting mixing effects can occur in the supersymmetric Higgs
sector if CP--noninvariant interactions are present. In the
decoupling regime these effects can become very large, leading to
interesting experimental consequences. Higgs formation in
$\gamma\gamma$ collisions with polarized beams proves particularly
interesting for observing such effects. However, exciting
experimental effects are also predicted in such scenarios  for $t
\bar{t}$ final--state analyses in decays of the heavy Higgs bosons
at LHC and in the $e^+e^-$ mode of linear colliders.

Detailed experimental simulations would be needed to estimate the
accuracy with which the asymmetries presented here can be
measured. Though not easy to measure, the large magnitude and the
rapid, significant variation of the CP--even and CP--odd
asymmetries through the resonance region with respect to both the
phase $\phi_A$ and the $\gamma\gamma$ energy would be a very
interesting effect to observe in any case.

\begin{acknowledgments}
The work of SYC was supported by the Korea Research Foundation Grant
(KRF--2002--041--C00081) and the work of JK by the KBN
Grant 2 P02B 040 24 (2003--2005) and
115/E--343/SPB/DESY/P--03/DWM517/2003--2005.
\end{acknowledgments}

\end{document}